# Tracing the temporal evolution of clusters in a financial stock market

Argimiro Arratia[a,1,*], Alejandra Cabaña[b,2]

[a]*Llenguatges i Sistemes Informàtics,*
*Universitat Politècnica de Catalunya, Barcelona, Spain*
[b]*Matemàtiques,*
*Universitat Autónoma de Barcelona, Barcelona, Spain*

**Abstract**

We propose a methodology for clustering financial time series of stocks' returns, and a graphical set-up to quantify and visualise the evolution of these clusters through time. The proposed graphical representation allows for the application of well known algorithms for solving classical combinatorial graph problems, which can be interpreted as problems relevant to portfolio design and investment strategies. We illustrate this graph representation of the evolution of clusters in time and its use on real data from the Madrid Stock Exchange market.

## 1. Introduction

The problem of classifying financial time series by some measure of similarity has received a lot of attention, although the emphasis has been mostly on what has been termed as *model–based clustering* (see [7], [8] and references therein). In this type of clustering the similarity of the financial time series

*Corresponding author

*Email addresses:* argimiro@lsi.upc.edu (Argimiro Arratia),
acabana@mat.uab.cat (Alejandra Cabaña)

[1]Research partially supported by Spanish Government MICINN projects: SESAAME TIN2008-06582-C03-02 and SINGACOM MTM2007-64007

[2]Research partially supported by Spanish Government MICINN projects: MTM2009-10893, and SESAAME TIN2008-06582-C03-02

is translated into the similarity of the models characterising them (usually an ARMA or GARCH process), and hence what is been classified is the dynamics of the variances or volatility of the series.

The present work can be framed into the category of *raw-data based clustering* [7], where the series are compared with respect to their history of prices, normally sampled at the same time interval. We are interested in identifying stocks whose return history strongly resembles each other in order to use this information for designing investment strategies, such as "pairs trading" [13].

Pairs trading is a stock trading strategy that attempts to capture the spread between two correlated stocks as they return to the mean price. Roughly speaking it consists on taking long and short positions on two stocks that tend to move together. The first step in this strategy is to find such a pair, but as correlation depends on the length of time span what the trader is really interested in is to identify short time correlations that are sustained through long periods of time, in order to capture the *correlation in movement*.

In this paper, we propose a graphical representation of the evolution in time of clusters of financial time series. The clustering is done through standard procedures with respect to a metric based on pair wise correlation coefficients. The temporal graph representation of the evolution of clusters through time is basically a weighted graph whose vertices are clusters of stocks, bound by a relation of similarity on their return history, and edges among clusters with non empty intersection are weighted by the cardinality of this intersection. The details of this construction are given in §2.

An added advantage of having this graph is that it allows to tackle certain questions about the behaviour of financial time series as classical problems on the combinatorics of graphs. To illustrate this last point, here is a couple of possible applications, which will be dealt with later:

1. Finding a path (or a fixed number of paths) of heaviest weight in the graph, translates to detecting the most stable clusters through time. In the context of finance it helps to detect those stocks whose return history mirror each other (but certainly at different scales) through long stretches of time.
2. Finding a *vertex cover* [4], translates into finding a (minimum) set of stocks that intersects all clusters in the temporal graph. This set of stocks can be considered as a group of representatives for the market index, in the sense of being the stocks that in conjunction best replicate the overall combined market price in a given period of time; or, from another point of view, it can be considered as a minimum portfolio covering the market.

We expand on these ideas in §3. Finally, in §4 we present an illustration of the proposed methodology of clustering, its graphical representation and combinatorial studies derived from it to real data from the Madrid Stock Exchange market.

## 2. Description of the temporal graph of clusters

This section outlines the algorithmic construction of the graph that describes the temporal evolution of clusters of a given set of financial time series, in a given time span. References for notation and facts on the statistical tools implemented, and in particular on financial time series can be found in the books by Tsay [12] and Brockwell and Davis [1].

*2.1. Adjusting to significant correlation*

We assume that daily returns for each of the stocks in financial markets are random variables. For each stock $X$, we observe $n$ values, considered as vectors $\mathbf{x} = (x_1, \ldots, x_n)$ corresponding to the *returns* during $n$ trading days; that is, each $x_k$ corresponds to *the variation of the daily closing price*:

$$x_k = \frac{P(k)}{P(k-1)} - 1 \qquad (1)$$

where $P(k)$ is the closing price of the stock at time instant $k$.

Given the observations $\mathbf{x}$, denoting $\bar{\mathbf{x}} = n^{-1} \sum_{i=1}^{n} x_i$ the sample mean, the sample autocovariance function is defined as $\hat{\gamma}(h) = n^{-1} \sum_{t=1}^{n-|h|}(x_{t+|h|} - \bar{\mathbf{x}})(x_t - \bar{\mathbf{x}})$ and the sample autocorrelation at lag $h$ is $\hat{\gamma}(h)/\hat{\gamma}(0)$ (see, for instance, [1]). Ljung–Box test can be applied in order to check whether the data are serially uncorrelated. Ljung–Box test is a so-called *portmanteau* test, in the sense that takes into account all sample autocorrelations up to a given lag. For $n$ time periods a good upper bound for the choice of lag is $\log(n)$ [1, 12].

As a first descriptive tool, we compute the sample correlation coefficients for pairs of stocks returns. For pairs of observations, $\mathbf{x}$ and $\mathbf{y}$, taken in a common period of $n$ time instants, the sample correlation coefficient is computed as

$$r(\mathbf{x}, \mathbf{y}) = \frac{\sum_{i=1}^{n}(x_i - \bar{\mathbf{x}})(y_i - \bar{\mathbf{y}})}{\sqrt{\sum_{i=1}^{n}(x_i - \bar{\mathbf{x}})^2 \sum_{i=1}^{n}(y_i - \bar{\mathbf{y}})^2}} \; .$$

This quantity $r = r(\mathbf{x}, \mathbf{y})$ is a consistent estimator of the true correlation $\rho$ between the random returns of the stocks being considered, which is a measure of the linear association between them. Correlation is only a measure of association, and has little use in prediction; however, it proves to be useful for detecting clusters of stocks that behave in association. It is often useful to test the *null hypothesis* $H_0 : \rho = 0$; that is, that the correlation coefficient $\rho$ equals zero. An appropriate statistic for this test is

$$t_0 = \frac{r\sqrt{n-2}}{\sqrt{1-r^2}}$$

which is distributed as a $t$ with $n - 2$ degrees of freedom under the null hypothesis $\rho = 0$. Thus, "big" values of $t_0$ (or $r$) will lead us to reject the null hypothesis. In fact these kind of data are rather log-normally distributed, but in practice, the distinction between simple and log-returns is not substantial [12]. Hence, we use the bounds given by the $t$ test critical points to determine significant correlations.

Table 1 shows the critical points $c_\alpha$ for the tests with critical region $|r| > c_\alpha$ for testing $H_0$, for levels $\alpha = 0.05$ and $0.01$ for different values of $n$, that is, the number of trading days considered. In other words, correlations equal or smaller than the critical point in absolute value will be statistically indistinguishable from zero at level $\alpha$ for the corresponding sample size $n$.

| $n$ | 5 | 10 | 20 | 40 | 60 | 240 |
|---|---|---|---|---|---|---|
| $c_{0.05}$ | 0.8783 | 0.6319 | 0.4438 | 0.3120 | 0.2542 | 0.1267 |
| $c_{0.01}$ | 0.9587 | 0.7646 | 0.5614 | 0.4026 | 0.3301 | 0.1660 |

Table 1: Critical points for the contrast for $H_0 : \rho = 0$ with critical region $\{|r| > c_\alpha\}$

2.2. *Clustering*

We reason as follows for defining the clusters in the most appropriate way. Any strategy for clustering a set $\mathcal{S}$ of data, under some notion of similarity, seeks to partition $\mathcal{S}$ in such a way that any two elements in the same subset (or cluster) are similar and two elements in different subsets are dissimilar. A sound procedure is to endow $\mathcal{S}$ with some metric $d$, related to the similarity criteria, and treat each cluster as a ball, in topological space $(\mathcal{S}, d)$, centered at some element and of a certain radius. There are many ways of partition $\mathcal{S}$ in balls, and these depend on the chosen center and radius.

For two financial returns series the similarity is measured by their correlation coefficient (with the appropriate adjustment as explained in the previous section). However, we are reluctant to use correlation values as a measure of proximity for a collection of returns since these do not constitute a metric. A second caveat is that once we choose a stock $A$ and find the subset $S_A$ of stocks with series of returns correlated to the series of returns $\mathbf{a}$ of $A$ observed in the same time period, the succeeding groups (or balls) have to be constructed from $\mathcal{S} \setminus S_A$, thus failing from considering all possible correlations of returns of stocks in $S_A$ for the successive stocks' returns fixed as center, and furthermore, the last stocks chosen will forcibly have few or none correlated match.

Therefore, in order to overcome this "order of selection" dependency plus the lack of metric to quantify the similarity distance, we do the following:

Let $\tau$ be a fixed time period.

- If it is the case that one wants to cluster stocks with positive correlated returns series, taken in the span of $\tau$, then for each stock $A$, with series of returns $\mathbf{a}$ in the time period $\tau$, we group into $\mathcal{G}_A$ all stocks $X$ whose series of returns $\mathbf{x}$, on the same time span of $\mathbf{a}$, have correlation with $\mathbf{a}$ higher than a positive $\delta$; that is,

$$X \in \mathcal{G}_A \iff r(\mathbf{a}, \mathbf{x}) > \delta.$$

The $\delta$ is determined by our statistical test explained in previous section (cf. Table 1) and depends on the sample size. For example, for 40 days period we take $\delta = 0.65$, a midpoint between 1 and the threshold of $c_{0.05} = 0.3120$, to ensure a significant correlation.

For the case of clustering stocks with negative correlated return time series, consider a negative $\delta$ and define $\mathcal{G}_A$ as $X \in \mathcal{G}_A \iff r(\mathbf{a}, \mathbf{x}) < \delta$.

- Next, for each pair of stocks $A$ and $B$ define

$$d(A, B) = 1 - \frac{|\mathcal{G}_A \cap \mathcal{G}_B|}{|\mathcal{G}_A \cup \mathcal{G}_B|}$$

(the term $\frac{|\mathcal{G}_A \cap \mathcal{G}_B|}{|\mathcal{G}_A \cup \mathcal{G}_B|}$ is known as Jaccard measure and $d(A, B)$ a Jaccard distance).

$d$ is a metric: it is positive, symmetric, $d(A, B) = 0$ iff $A = B$, and verifies triangle inequality. The last property follows by straightforward computation:

$$\frac{|\mathcal{G}_A \cup \mathcal{G}_B \setminus \mathcal{G}_A \cap \mathcal{G}_B|}{|\mathcal{G}_A \cup \mathcal{G}_B|} + \frac{|\mathcal{G}_B \cup \mathcal{G}_C \setminus \mathcal{G}_B \cap \mathcal{G}_C|}{|\mathcal{G}_B \cup \mathcal{G}_C|} \geq$$
$$\frac{|\mathcal{G}_A \cup \mathcal{G}_B \setminus \mathcal{G}_A \cap \mathcal{G}_B| + |\mathcal{G}_B \cup \mathcal{G}_C \setminus \mathcal{G}_B \cap \mathcal{G}_C|}{|\mathcal{G}_A \cup \mathcal{G}_B \cup \mathcal{G}_C|} =$$
$$1 - \frac{|\mathcal{G}_A \cap \mathcal{G}_B \cap \mathcal{G}_C|}{|\mathcal{G}_A \cup \mathcal{G}_B \cup \mathcal{G}_C|} \geq 1 - \frac{|\mathcal{G}_A \cap \mathcal{G}_C|}{|\mathcal{G}_A \cup \mathcal{G}_C|}.$$

One can see that if $A$ and $B$ are similar, in the sense that their respective groups $\mathcal{G}_A$ and $\mathcal{G}_B$ have almost the same members, then $d(A, B)$ is close to 0; whilst if $\mathcal{G}_A$ and $\mathcal{G}_B$ greatly differ then $d(A, B)$ is close to 1. Note that $d$ is a stronger measure of similarity than taking a direct correlation coefficient, since it says that *two stocks are similar, not only if they are correlated but also correlated to almost all the same stocks.*

- Apply a clustering method based on the distance metric $d$. The method we have chosen is the *hierarchical clustering*, which is an agglomerative algorithm that stepwise builds clusters, beginning from singletons and successively increasing their size by merging elements that are nearer according to $d$ (see [6]). The decision of the final number of clusters that are to be considered is made by the standard procedure of cutting at mid level the tree representation, or dendogram, of the hierarchical clustering. After the cut, the clusters that contain only one element are merged into one group and labelled as outliers (elements whose returns are not significantly correlated to any other).

Other methods of clustering can be applied leading to different experimental results. For example, a most popular choice is the $k$–means algorithm [6]. However with this algorithm the user must arbitrarily choose before hand the number of clusters, or rather use a test like *gap statistics* [11] to make a more accurate selection of the number of clusters. In any case, with $k$–means every element is assigned to some cluster to the expense of lowering the similarity measure within clusters. Thus, depending on the application in mind the user should use one method or the other. As mentioned in the Introduction one of our concerns is identifying groups of stocks with high similar return behaviour in order to apply diverse investment strategies (e.g. pairs trading). Hierarchical clustering serves well this purpose. On the other hand, the method of $k$-means might serve the purpose of producing clusters corresponding to the different industrial sectors of the market, which is a most common application of clustering of financial time series under the model–based paradigm (see e.g. [8]).

We repeat the computations for further temporal intervals, and use combinatorial graphs algorithms to build a (weighted) representation of the evolution of clusters in time. This construction is explained in the next section.

*2.3. Temporal graph of clusters*

Let $\mathcal{S}$ be a set of market stocks. Formally, $\mathcal{S}$ is a set of financial time series corresponding to these stocks. Let $T$ be the time period selected for analysis, which is partitioned into $m$ successive sub-periods of time $\tau_1$, $\tau_2$, ..., $\tau_m$. For example, in our experiments presented in §4 we take $T$ to be two consecutive years and each $\tau_i$ two successive months in the span of $T$. For each $i = 1, \ldots, m$, let $\mathcal{C}_i$ be the collection of clusters obtained in each temporal interval $\tau_i$, constructed as describe in §2.2, and assuming we applied hierarchical clustering, and let $n_i = |\mathcal{C}_i|$. Let $S_{i,j}$, $1 \leq i \leq m$, $1 \leq j \leq n_i$, be the clusters in $\mathcal{C}_i$ with at least two elements, and $Q_i$ be the subset of elements with no significant correlation in time segment $\tau_i$, for $1 \leq i \leq m$.

We define a directed graph $\mathcal{G}$ with vertex set the collection of subsets

$$\{S_{i,j} : 1 \leq i \leq m, 1 \leq j \leq n_i\} \cup \{Q_i : 1 \leq i \leq m\}$$

and weighted edge set

$$\{(S_{i,j}, S_{i+1,k}, |S_{i,j} \cap S_{i+1,k}|) : 1 \leq i \leq m-1, 1 \leq j \leq n_i, 1 \leq k \leq n_{i+1}\}$$

that is, we link a cluster $S_{i,j}$ in the time segment $\tau_i$ with a cluster $S_{i+1,k}$ in the next time segment $\tau_{i+1}$ as long as their intersection is non empty, and the cardinality of their intersection is the weight assigned to the link. The sets $Q_i$ remain as isolated vertices. Beware that these $Q_i$'s do not represent stocks whose returns have some significant (negative or not) correlations with the returns of other stocks. These are just variables that do not pass our statistical test and hence can not be considered significantly correlated. We call $\mathcal{G}$ the *Temporal Graph of Clusters* (TGC) for the set of stocks $\mathcal{S}$ in the time segmentation $T = \bigcup_{1 \leq i \leq m} \tau_i$.

By considering the cardinality of the intersection of the clusters as the weight for the edges of $\mathcal{G}$ we seek to capture the most persistent or stable clusters through time. These would be located on the paths of heaviest weight (considering the weight of a path as the sum of the weights of the edges that comprise it).

*2.4. The Algorithm*

Let $\mathcal{S}$ be a set of financial time series corresponding to some market stocks. Let $T$ be the time period selected for analysis, which is partitioned into $m$ successive sub-periods of time $\tau_1, \tau_2, \ldots, \tau_m$.

**Pre–processing:** Depending on the provider of the data there will be some pre–processing to be done. For instance, if the financial data is retrieved from *yahoo.com*, the file comes in columns labelled *Date*, *Open*, *High*, *Low*, *Close* and *Volume*. Then one has to produce, for each stock, a two columns table of *Date* and *Return* (equation 1). Denote by $\hat{\mathcal{S}}$ the resulting pre-processed data $\mathcal{S}$.

**Input:** $\hat{\mathcal{S}}$, a list of tables (with entries *Date* and *Return*) containing the return history of each stock in the time period $T$. The partition of $T = \{\tau_1, \ldots, \tau_m\}$.

**Part I** For each time segment $\tau_i$, $1 \leq i \leq m$, determined by its initial date and ending date, do:

1. Build the matrix $R$ of series of returns for the period $\tau_i$ of all stocks in $\hat{\mathcal{S}}$. $R$ is the concatenation as columns of each of these return series taken on the same period $\tau_i$.
2. Compute clusters following the method explained in §2.2. This part consist of two steps.

    **step 1** For each column of $R$, containing the return series of some stock $A$ in the time period $\tau_i$, compute the set $\mathcal{G}_A$ of all other stock $X$ whose return series has correlation coefficient with that of $A$ above the threshold obtained from Table 1 (respectively, below the threshold if what is desired to measure is negative correlation). Then define the *dissimilarity* matrix $M$ where each entry is the value of the jaccard distance $d(A, B)$, defined in §2.2.

**step 2** Apply the hierarchical clustering algorithm to the matrix $M$. The output is a list of clusters which is formatted as a column vector $VCol$.

3. Define the vertex set $V$ as the successive concatenations of column vectors $VCol$.

**Part II**
1. Remove from $V$ all those entries (clusters) containing a single element, and rename the result $V'$.
2. Compute the edge relation $E$ on $V'$ by linking an element of column $i$ in $V'$ to an element in column $i+1$, whenever the intersection is non zero. The weight of the edge is the cardinality of the intersection.

**Output:** $V$ and $E$.

## 3. Graph combinatorics applied to financial analytics

Throughout this section $\mathcal{G}$ is the weighted graph representation of the temporal evolution of clusters in a financial market, the TGC as constructed in §2.3, without the isolated vertices $Q_i$. Applying graph combinatorial techniques to $\mathcal{G}$ we can solve problems relevant to financial analytics. We shall address three problems:

1. *Stable clusters*, consist on finding the clusters (or sub-clusters) of stocks that appear more frequently in consecutive intervals of time. On the TGC $\mathcal{G}$ this amounts to finding the heaviest paths from the first time interval to the last. The problem of deciding the heaviest weighted path in a graph is in general NP-hard; however, on acyclic directed graphs polynomial time solutions are known using dynamic programming ([4] problem [ND29]). We shall make use of such algorithmic methodology for producing our own solution to this problem on $\mathcal{G}$. The relevance of this graph problem to financial analytics is of a large extent. For example, as suggested in the Introduction it helps detect those stocks that correlate through different by continuous time intervals, and hence become candidates for a pairs trading investment strategy [13].
2. *The trace of a given stock*, consist on finding the sequence of clusters through consecutive intervals of time where a given stock is contained in all. On $\mathcal{G}$ this is simply to find a path from the first time interval to the last, subject to the restriction that a given stock must appear in all vertices (clusters) conforming the path. This allows to identify the market associations of a given stock through time.
3. *Stock cover*, consist on finding the smallest possible set of stocks that intersects every (non singleton) cluster in $\mathcal{G}$. This is the Hitting Set problem ([4] problem [SP8]), which is a form of the Vertex Cover problem on graphs, both NP-hard. Hence, the best we can aim for is an approximate solution, which we construct following a greedy strategy. The interest of this computational query is to discover those stocks that essentially represent the market behaviour through a sequence of time periods; thus, conforming a portfolio that *covers* the market.

In the next three sections we give the algorithmic details. As a reference for the programming schemes and data structure that we implement in our algorithms we recommend [2].

*3.1. Stable clusters*

Given a positive integer $k \geq 1$ and TGC $\mathcal{G} = \langle V, E \rangle$, with $V = \{S_{i,j} : 1 \leq i \leq m, 1 \leq j \leq n_i\}$, we want to find the $k$ first heaviest paths starting at any cluster $S_{i,j}$ with no ingoing edge.

Let $\{\tau_1, \ldots, \tau_m\}$ be the sequence of time segments which partition the total time period $T$. We view the edges in $\mathcal{G}$ as directed going from time segment $\tau_i$ to time segment $\tau_{i+1}$. Then we add an extra (dummy) cluster $S_{m+1,1}$ after time period $\tau_m$, and which serves as a sink in the sense that every cluster $S \in V$ that do not have an outgoing edge (i.e. is terminal) is connected by an edge of weight 1 to $S_{m+1,1}$. Now, a brute force solution is to find all paths that end in $S_{m+1,1}$ (say by Breadth First Search), order them by weight and keep the $k$ heaviest. This, however, would incur in exponentially many comparisons, as we can have a worst case containing $O(n^m)$ many paths, where $n$ is the maximum number of clusters per time segment and $m$ is the number of time segments. Instead we apply a bottom-up approach (aka dynamic programming) where at step $i$, we look at time segment $\tau_i$ and define, for each cluster $S_{i,j}$, a heap (i.e. a tree with nodes having a positive value or *key*) where the root, labelled $S_{i,j}$, has key 0, its descendent will be all heaps defined in previous time segments $\tau_{i'}$, $i' < i$, for which there is an edge from $S_{i',j'}$ to $S_{i,j}$, and the keys of all *leaves* of these descendent heaps are updated by adding the weight of the edge $(S_{i',j'}, S_{i,j})$, denoted $w(S_{i',j'}, S_{i,j})$.[3]

---

[3]By a leaf of a heap we understand a node with no ingoing edges. By updating the

Observe that the total weight of a path in the heap of root $S_{i,j}$ is the key of the leaf or initial node of such path. We order these leaves by key value and keep the paths for the $k$ leaves with highest key, removing the paths of the remaining leaves before proceeding to step $i+1$ of algorithm.

Below we present our algorithmic solution written in pseudo–code style. We use the following abbreviations for functions whose implementation we do not specify but are obvious: for a vertex $S$, $key[S]$ keeps the key value of $S$ in a heap which should be clear from context; for a heap $h$, $leaves(h)$ is the set of leaves of heap $h$; for two heaps $h$ and $h'$, $append(h', h)$, returns a heap made by drawing an edge from the root of heap $h'$ to the root of heap $h$.

```
Algorithm:   k first heaviest paths
1.     input:   k > 0, V = {S_{i,j} : 1 ≤ i ≤ m, 1 ≤ j ≤ n_i}, E
2.     (preprocessing) Add a sink S_{m+1,1}:
3.         for each S ∈ V with no outgoing edge do
4.             define new edge (S, S_{m+1,1}) of weight 1;
5.         end for;
6.     for each i ∈ {1,...,m+1} do
7.         for each j ∈ {1,...,n_i} do
8.             define heap h_{i,j} with unique element
9.             labelled S_{i,j} and key[S_{i,j}] = 0;
10.            for each S_{i',j'} such that (S_{i',j'}, S_{i,j}) ∈ E do
11.                h_{i,j} = append(h_{i',j'}, h_{i,j});
12.                for each S ∈ leaves(h_{i',j'}) do
13.                    key[S] = key[S] + w(S_{i',j'}, S_{i,j});
14.                end for;
15.            end for;
16.            remove from h_{i,j} all but k paths to S_{i,j}
17.            of heaviest weight; // The weight of each path
18.                // can be read-off from the key of each leaf.
19.        end for;
20.    end for;
21.    output:   h_{m+1,1};
```

---

keys of the leaves only we significantly reduce computation time since we do not need to traverse all the heap. Also the information about the key value of inner nodes is irrelevant for our purposes. For further details on heaps see [2]

One can see that for $m$ time segments and a maximum of $n$ many clusters per time segment, the algorithm has a worst-case running time bound of $O(mn^3)$, which is much better than the naive solution explained at the beginning.

*3.2. The trace of a given stock*

Given an specific stock (a company's name) and the TGC $\mathcal{G} = \langle V, E \rangle$, we want to find the path in $\mathcal{G}$ comprise of clusters containing the stock. This is a straightforward path finding algorithm (e.g. breath first search) with the restriction that at each step we check that the given stock is in the currently visited cluster.

*3.3. Stock cover*

Given a TGC $\mathcal{G} = \langle V, E \rangle$, with $V = \{S_{i,j} : 1 \leq i \leq m, 1 \leq j \leq n_i\}$, we want to find a (minimum) set of stocks that intersects every (non singleton) cluster in $\mathcal{G}$.

We apply the following greedy strategy: pick the stock that shows up more often in the collection of clusters $V$; then remove the clusters containing this stock (i.e. covered by the stock). Repeat with remaining clusters.

In the pseudo-code implementation of the above strategy we define an *incidence* matrix $\mathcal{M}$ for the set of clusters $V$, where rows are labelled by the companies identifier or *ticker* and columns labelled by the clusters. $\Pi$ represents the set of tickers (which is given as input). Then there is a 1 in entry $\mathcal{M}(\pi, S)$, $\pi \in \Pi$ and $S \in V$, if stock represented by ticker $\pi$ is in cluster $S$, or a 0 otherwise. By $dim(\mathcal{M})$ we denote the dimension of matrix $\mathcal{M}$, which is the number of rows multiplied by the number of columns. Details follows.

```
Algorithm:  stock cover
1.    input:   V = {S_{i,j} : 1 ≤ i ≤ m, 1 ≤ j ≤ n_i},  Π
2.    (initialize) Γ = ∅;
3.    Define a matrix M with rows labelled by each ticker
      π ∈ Π and columns labelled by each cluster S ∈ V,
      and there is a 1 in entry M(π, S), if stock represented
      by π is in cluster S, or a 0 otherwise;
4.    while dim(M) ≠ 0 do //dim(M) = (num. rows) × (num. columns)
5.       order the rows decreasingly with respect to
```

```
            the number of 1 appearing in it;
            // Thus the first row corresponds to the stock
            // appearing more often in the clusters.
6.          let π₁ be the ticker of the stock labelling the first row;
7.          Γ = Γ ∪ {π₁};
8.          delete every column with label S ∈ V for which M(π₁,S) = 1;
9.          delete row labelled π₁;
10.         rename M the reduced matrix;
11.     end while;
12.     output:  Γ;
```

The running time of this algorithm is bounded by $dim(\mathcal{M})$; that is by $|\Pi| \cdot |V|$. More importantly is to see that in any case the size of the solution is not too big with respect to the size of the full set of companies involved. We show that the size of the solution set $\Gamma$ is at most one half of the size of the set of companies $\Pi$.

**Proposition 1.** $|\Gamma| \leq \frac{1}{2}|\Pi|$ .

Proof. Each cluster contains at least two stocks. The worst situation is that any stock appears in at most one cluster. The algorithm selects a stock $\pi_1$ (of highest weight) and removes the cluster where it belongs together with at least another stock that will not enter the solution set (otherwise its weight is higher than $\pi_1$'s and would have been selected before). Thus, at most 1/2 of the stocks enters $\Gamma$. □

### 4. An application: Exploring the Spanish market

In this section we describe the results obtained by applying our algorithms on real data from the Madrid Stock Exchange during the years 2008 and 2009. We have chosen 34 of the big cap companies contained in the main index of this market, IBEX 35[4]. The IBEX 35 is the benchmark index composed of the 35 most liquid securities listed on the Madrid Stock Exchange. It is a market capitalization weighted index, calculated by a recursive formula [10].

---

[4]IBEX 35 is an index composed by 35 companies, subject to changes in its composition. In the time span considered we found these 34 as the largest set of persistent companies.

Thus, IBEX 35 is a geometric index (as opposed, for example, to Dow Jones which is arithmetic), with a value at a time instant highly correlated with the values of its components of higher market capitalization. Hence, according to the formula, an investor that wants to have a portfolio that replicates this index should buy the shares of the "biggest of the big cap" companies. Nonetheless, we shall show that other options are possible.

The companies we analyse are the following (the symbol in parenthesis is the market identifier or *ticker*, which we use in our graphical representation): Abertis (ABE), Abengoa (ABG), Actividades de Construcción y Servicios (ACS), Acerinox (ACX), Acciona (ANA), Bankinter (BKT), Banco Bilbao Vizcaya Argentaria (BBVA ), Bolsas y Mercados Españoles (BME), Banesto (BTO), Criteria Caixa Corp. (CRI), Endesa (ELE), Enagas (ENG), Fomento de Construcciones y Contratas (FCC), Ferrovial (FER), Gamesa (GAM), Gas Natural (GAS), Grifols (GRF), Iberdrola (IBE), Iberia (IBLA), Iberdrola Renovables (IBR), Indra Sistemas (IDR), Inditex (ITX), Mapfre (MAP), Banco Popular (POP), Red Eléctrica (REE), Repsol YPF (REP), Banco Sabadell (SAB), Banco Santander (SAN), Arcelor Mittal (MTS), Sacyr Vallehermoso (SYV), Telefónica (TEF), Telecinco (TL5), Técnicas Reunidas (TRE), Obrascón Huarte Lain (OHL). This set of companies will be referred to as *Ibex Big Cap* (or *IbexBC* for short).

The data have been obtained from *Yahoo Finance*, and it is stored and managed with MySQL. The algorithms (see §2.4 and §3) were programmed in R, making use of some of the packages contained in this statistical software [9]. Although we test our clustering algorithm and graphical representation on this particular market, they are applicable to any financial stock market.

Before computing clusters we test for serial autocorrelation of the 34 series, as indicated in §2.1. We have performed the Ljung–Box test for all the above assets by means of the instruction `Box.test(x,lag=6,type="Ljung")` in R. Applying Ljung–Box test up to lag $6 \approx \log(256)$, for daily data, Table 2 shows the $p$-values obtained for IBEX companies in the years 2008 and 2009. The boldface numbers correspond to $p$ values smaller than 0.05. There is no statistically significant dependence in most assets returns series in the period 2009. There are few more companies that show serial autocorrelation in their prices during 2008.

We apply the algorithm as described in §2.4, with time period $T$ from 1-06-2008 to 1-08-2009, segmented bimonthly, thus $T$ is partitioned into 7

| Ticker | 2008 $p$-value | 2009 $p$-value | Ticker | 2008 $p$-value | 2009 $p$-value |
|---|---|---|---|---|---|
| ABE  | 0.1087     | 0.4052     | IBE  | **0.0055**  | 0.3469     |
| ABG  | 0.2771     | 0.8080     | IBLA | 0.1525      | 0.9290     |
| ACS  | 0.4305     | 0.9764     | IBR  | **0.0004**  | 0.3875     |
| ACX  | **0.0271** | 0.1748     | IDR  | **0.0051**  | 0.3095     |
| ANA  | 0.8196     | 0.4961     | ITX  | **0.0039**  | 0.1386     |
| BKT  | 0.5887     | 0.4525     | MAP  | **0.0016**  | 0.1612     |
| BBVA | **0.0159** | **0.0124** | POP  | **0.0329**  | 0.1492     |
| BME  | 0.5205     | **0.0185** | REE  | **0.0004**  | 0.8557     |
| BTO  | **6 e-06** | 0.2817     | REP  | **0.0112**  | 0.9884     |
| CRI  | **0.0369** | 0.0517     | SAB  | 0.4162      | **0.0043** |
| ELE  | 0.5658     | 0.7748     | SAN  | 0.1390      | 0.6888     |
| ENG  | 0.1697     | 0.9862     | MTS  | **0.0114**  | 0.9009     |
| FCC  | 0.3361     | 0.5496     | SYV  | 0.1213      | 0.2185     |
| FER  | 1          | **0.001**  | TEF  | **2e-06**   | 0.5921     |
| GAM  | **0.038**  | 0.9006     | TL5  | 0.5294      | 0.5735     |
| GAS  | **0.0094** | **0.0141** | TRE  | 0.2801      | 0.4134     |
| GRF  | **0.0165** | 0.1639     | OHL  | 0.4200      | **0.0295** |

Table 2: $p$-values for $IbexBC$ components in 2008 and 2009

periods, $\tau_1, \ldots, \tau_7$.

Figure 1[5] presents the correlation matrix for the first period of our sample, 1-06-2008 to 1-08-2008 (the scale on the right represents the correlation values found). Next, Figure 2 is the dendogram representation of the result of applying the hierarchical clustering algorithm to the dissimilarity matrix associated to the correlation matrix. Figure 3 is the graph of clusters formed through the different time periods in the selected years: the green boxes (boxes in first row from bottom) contain the time periods for each return series; the pink boxes (boxes in second row from bottom) collect companies with correlations below our established threshold (hence, no conclusion can be drawn for these); and the blue boxes (all remaining boxes above second row from bottom) represent non singleton clusters, which also contain a label $ij$ to indicate its position in the adjacency matrix representation, so for example cluster 13 is {ACX, ANA, SYV}; the edges are weighted by the intersection measure.

Observe that in the time period $\tau_3 = [01/10/2008, 01/12/2008]$ is where the largest cluster appears (cluster 31). This time frame corresponds to a bullish period when all prices were rising on the market and thus align in correlation as expected. Also there is one title, GRF, that does not belong to any cluster in any time period. Incidentally the business of the corresponding company, Grifols, is far from the industrial sectors to which all other companies of *IbexBC* belong. Notice also that the resulting clustering is not sectorial. This is not surprising as we are comparing time series with respect to their first order moments, disregarding volatility which has been reported to discriminate industrial sectors (see, e.g. [8]).

The procedure for building the temporal graph of clusters for negatively correlated stocks is the same as in the positive case but considering as good correlation values those *below* a negative threshold given by our critical points shown in Table 1. For our particular subject of experimenting our clustering method, namely *IbexBC*, we found that there is no time segmentation for the span of the years 2008 to 2009 that will produce clusters of stocks with significant negative correlations. On periods of time of 40 or 60 days of returns there are very few stocks that have negative correlations below -0.2 (see for example Figure 1), which is not a significant value. For four or six months long time segments the lowest correlation coefficient we got was about -0.25. For the full year 2008 the minimum correlation was -0.02, and for the year 2009, -0.22. None of these values can be considered as significant negative correlations, according to our criteria from §2.1.

---

[5]All figures are in the Appendix.

*4.1. Graph combinatorics*

1. Applying the *k first heaviest paths* algorithm to *IbexBC* for the years 2008-2009 and $k = 3$, we found the following three heaviest paths (the number on top of the arrow is the weight of the corresponding edge):

$$14 \stackrel{5}{\mapsto} 21 \stackrel{6}{\mapsto} 31 \stackrel{6}{\mapsto} 44 \stackrel{3}{\mapsto} 52 \stackrel{3}{\mapsto} 62 \stackrel{4}{\mapsto} 72$$

$$14 \stackrel{2}{\mapsto} 23 \stackrel{4}{\mapsto} 31 \stackrel{6}{\mapsto} 44 \stackrel{3}{\mapsto} 52 \stackrel{3}{\mapsto} 62 \stackrel{4}{\mapsto} 72$$

$$14 \stackrel{5}{\mapsto} 21 \stackrel{6}{\mapsto} 31 \stackrel{3}{\mapsto} 41 \stackrel{2}{\mapsto} 54 \stackrel{1}{\mapsto} 62 \stackrel{4}{\mapsto} 72$$

On these heaviest paths we find the subsets of companies with higher tendency to correlate through time (correlation in movement). We find that these belong mostly to three sectors: banks, energy and construction. On the second heaviest path we found the two most correlated pair, namely, the two banks SAN and BBVA, thus making a good candidate for a *pairs trading* strategy. To find more pairs to which apply the pairs trading we look for paths where the edge weight is 2 and remain constant for consecutive periods of time. This is the case for the path $16 \mapsto 24 \mapsto 33$, with all clusters consisting of ENG and REE, two companies from the energy sector. Interestingly these two companies remain in the boxes of unclassified (pink boxes) of the graph through all 2009.

2. Applying the *stock cover* algorithm to *IbexBC* for the years 2008-2009 we found the following cover: {BBVA, FCC, REP, REE, TRE, ACX, ABG, CRI, TEF, MAP, BME, ANA, MTS, TL5}

These fourteen companies represent a minimum set of stocks that intersects all non isolated clusters in the TGC for *IbexBC* through 2008 and 2009. It constitute a portfolio that would have replicated the behaviour of the market index IBEX 35 throughout those years. It is not unique (for example, BBVA can be substituted by SAN), and its composition is a consequence of the order of selection imposed by the algorithm. Another issue is that for the sake of completeness we should add to the cover found by the algorithm those stocks that are not in any cluster, such is the case of GRF.

## 5. Conclusions

We have proposed a graphical tool in order to monitor the temporal evolution of clusters of financial time series; that is, a representation of the *clusters in movement*. We have exploited some useful links between graph combinatorics and financial applications, which shows how problems in the former field translates to problems in the latter and used some known combinatorial methods from graph theory to produce sound answers for problems about financial markets.

The finding of the heaviest paths in our model confirmed a popular observation known by many local brokers that the banks BBVA and SAN conform the most stable cluster through any period of time, and further we see that this duo tends to participate in the largest clusters of IBEX 35 components at different periods of time; hence being both together a driving force of the Spanish market. Additionally, our experiments revealed that pairs of stocks that remain clustered together through long periods of time tend to belong to the same industrial sectors, as is the case of BBVA and SAN (banks), FCC and FER (construction), or ENG and REE (energy). These pairs can be good candidates for applying a pairs trading strategy.

The stock cover algorithm produces an approximate solution with at most half of the stocks available, in the worst scenario. The optimal solution is not algorithmically feasible and will certainly depend on the length and the number of time segments considered in the study. The approximate solution can be taken as a starting group of stocks that can be shaped up to obtain an *adequate* portfolio (according to Benjamin Graham [5, ch. 5] "a minimum of ten issues and a maximum of about thirty" is adequate). However, adequacy is not only a matter of size, but among other issues is about diversification (read [5]). From our experiments with negative correlation we conclude that the big cap companies of IBEX 35 are, in general, always positively correlated. Thus, an investor can not have a balanced portfolio consisting only on the big cap companies in the Madrid Stock Exchange.

# Appendix: Figures

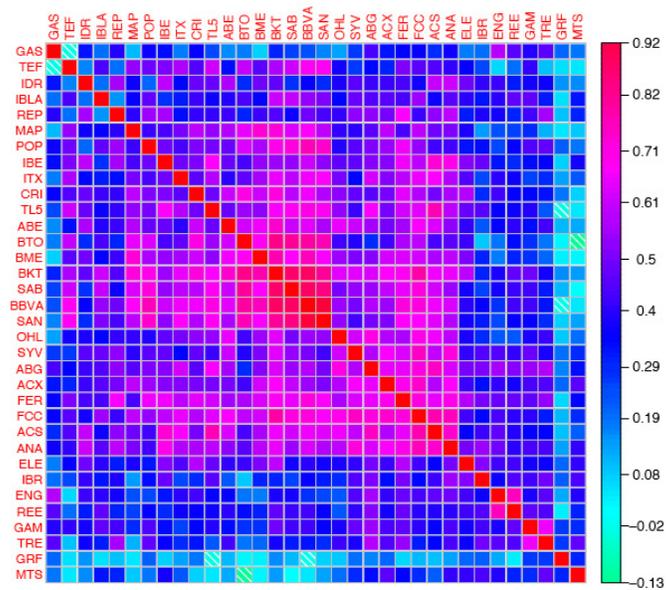

Figure 1: Correlation matrix for time segment 1-6-2008 to 1-8-2008

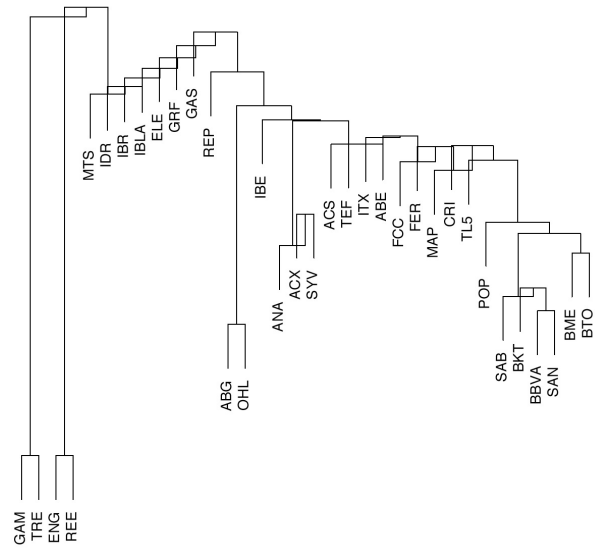

Figure 2: Dendogram for the correlation matrix of Figure 1

Figure 3: Temporal Cluster Graph for IBEX 35, years 2008-2009